# Conceptual Design and Preliminary Results of a VR-based Radiation Safety Training System for Interventional Radiologists

Yi Guo[1], Li Mao[2], Gongsen Zhang[1], Zhi Chen[1], Xi Pei[1,*], X. George Xu[1,2,*]

[1] Department of Engineering and Applied Physics, University of Science and Technology of China, Hefei, Anhui 230026, China

[2] Nuclear Engineering Program, Rensselaer Polytechnic Institute, Troy, NY 12180, USA

**Abstract.** Recent studies have reported an increased risk of developing brain and neck tumors, as well as cataracts, in practitioners in interventional radiology (IR). Occupational radiation protection in IR has been a top concern for regulatory agencies and professional societies. To help minimize occupational radiation exposure in IR, we conceptualized a virtual reality (VR) based radiation safety training system to help operators understand complex radiation fields and to avoid high radiation areas through game-like interactive simulations. The preliminary development of the system has yielded results suggesting that the training system can calculate and report the radiation exposure after each training session based on a database precalculated from computational phantoms and Monte Carlo simulations and the position information provided in real-time by the MS Hololens headset worn by trainee. In addition, real-time dose rate and cumulative dose will be displayed to the trainee by MS Hololens to help them adjust their practice. This paper presents the conceptual design of the overall hardware and software design, as well as preliminary results to combine MS HoloLens headset and complex 3D X-ray field spatial distribution data to create a mixed reality environment for safety training purpose in IR.

## INTRODUCTION

Each year over 10 million fluoroscopically guided interventional (FGI) procedures are performed for diagnostic or therapeutic purposes [1]. Physicians, technologists, and nurses are inevitably exposed to a high level of ionization radiation during FGI procedures due to scattered X-ray from the patient. The cumulative dose over an operator's career can be substantial especially with the increasing workload as a result of the rapidly maturing field. An alarming number of brain and neck tumors, as well as cataracts, have been reported for FGI operators who are found to have insufficient training of radiation physics and safety [2-8]. Occupational radiation protection during FGI procedures has been a top concern for regulatory agencies, professional societies, and radiologists [9-13]. ICRP has recently recommended a drastically more restrictive annual dose limit of 20 mSv for the lens of the eye from the previous limit of 150 mSv [14]. In 2011, about 40% of FGI physicians from a high-volume medical institution in the U.S. received an estimated eye lens dose equal or greater than the new ICRP limit [15]. There is an urgent need for research about radiation protection of FGI operators [16].

Previous studies showed that radiation doses to the eye lens and brain depended heavily on operator's posture and protective equipment [11,15]. It is worth noting that the X-ray field is invisible to the interventionist who is occupied with the surgical procedure, thus often failing to minimize radiation exposure. Literature surveys show that: (1) Factors contributing to operator radiation exposure are not well quantified, (2) The few available computational studies were based on human models that are not anatomically realistic, (3) There is no computer-based dose-feedback training system to help FGI operators to practice radiation safety.

This paper describes the conceptual design and preliminary development of a virtual-reality (VR) based X-ray safety training system for radiologists and nurses in FGI using Microsoft HoloLens glasses and precalculated radiation field data base. A set of deformable three-dimensional human anatomical phantoms and a Monte Carlo radiation dose computing code were used to generate a database of X-ray field under various exposure scenarios. The X-ray field data were then incorporated into MS Hololens to provide real-time feedback of dose rate and cumulative dose according to the position information of the trainee.

## METHOD

### The framework of the VR training system

The framework of the VR training system is shown in Figure 1. The system consists mainly of three components: the PC, Mixed Reality headset MS Hololens, and the trainee. The software on the PC provides functions including graphical user interfaces, radiation field database and dose reporting, training management, and learning assessment. Figure 2 shows the interfaces of the training management system. For each training session, the administrator will specify parameters that are related to the radiation exposure,

*Yi Guo and Li Mao contributed equally to this work

including X-ray tube voltage, filtration, field of view, KAP per minute. These parameters will be used to fetch the corresponding radiation field data from the precalculated database. During training, the administrator will turn on the real-time visual feedback assistance to help trainee adjust their practice. The real-time visual feedback feature of the training system is designed to display the radiation field to the trainee via see-through Hololens, and provides visible dose rate and cumulative dose readings to the trainee in real-time. After each training session, the training system can calculate and report the radiation exposure based on the radiation field data and position information provided by the Hololens headset worn by trainee during the training. The training data of each trainee is managed by the training management system. In addition, the training system also provides support for learning assessment which allows the evaluation of the effectiveness of the training system.

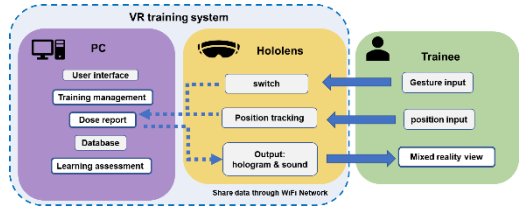

Figure 1. Framework of the VR training system

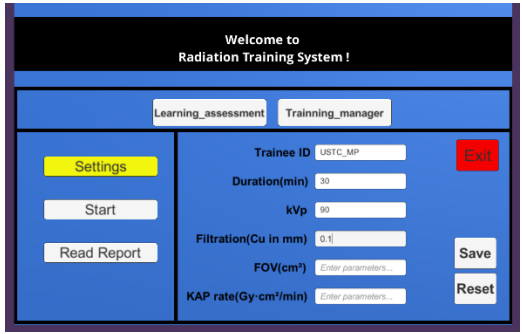

Figure 2. The preliminary graphical user interfaces of the VR training management system

**Computational human phantoms**

In order to accurately calculate the radiation fields and organ doses in the FGI suite, we used deformable whole-body computational human phantoms to simulate the patient who are exposed to X-rays during the FGI procedure. Existing computational phantoms can be classified into three categories: stylized phantoms, voxel phantoms and boundary representation (BREP) phantoms [17]. Among them, BREP phantoms are more suitable for deformation and adjustment. One can easily perform a series of geometric operations on BREP phantoms, including stretching, chamfering, mixing, dislocation, peeling and torsion. In this study, we used deformable RPI-Adult Male and Female [18] BREP phantoms for radiation field calculations. Figure 3 shows RPI phantoms (Figure 3a), as well as 3D rendering of the FGI suite including an operator and a patient (Figure 3b).

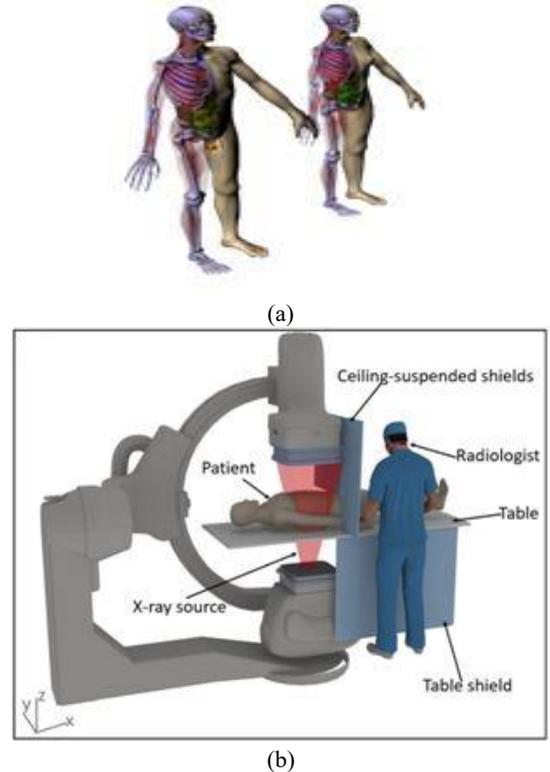

Figure 3. Deformable computational phantoms used to perform Monte Carlo simulations of radiation field. (a) RPI-Adult Male and Female phantoms. (b) The radiation field simulation showing 3D rendering of the FGI suite.

**Radiation data calculation**

Using deformed phantoms described earlier and the Monte Carlo methods [19], we created a database of the X-ray field data under different exposure scenarios in FGIs.

To obtain the radiation field in FGI suite, we calculated the dose distribution on a rectangular grid overlaid on top of the FGI suite model, with 5 cm × 5 cm × 5 cm cube as mesh cell. The obtained dose is averaged over each mesh cell. In the simulation, X-ray is emitted from an X-ray source below the patient's body and detected by a detector above the patient, as shown in the Figure 4. We calculated the radiation field under different tube voltages (80 kVp, 90 kVp) and different X-ray field of view (FOV)(10 × 10 cm$^2$, 20 × 20 cm$^2$, 30 × 30 cm$^2$, 40 × 40 cm$^2$).

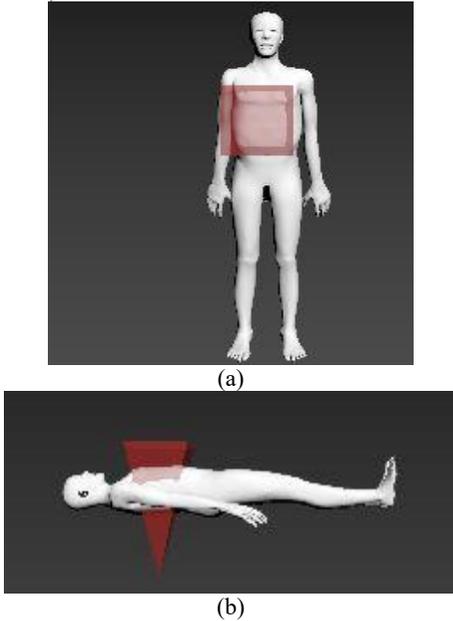

(a)

(b)

Figure 4. Patient phantom and X-ray under different views. (a) Top view. (b) Side view.

**Visualization of the radiation field**

After obtaining the distribution of the X-ray field, the data needs to be processed and converted into a format that can be read by the VR program. At the same time, couch, patient and operators also need to be modeled to simulate the real interventional procedures. We developed a visual program based on OpenGL to display the radiation field and save it in FBX format. The radiation field data are converted into text files, and then processed by the developed program. The color coding of the radiation field is according to the radiation intensity of each mesh cell. Different colors represent regions with different radiation level - red represents high radiation area and blue represents low radiation area. After completing the radiation field modeling, the operators, couch, patient and X-ray source are modeled respectively by using 3D Studio Max and Unity3D, and then saved in FBX format.

**Visual real-time feedback**

To help the trainee better understand the radiation field, the training system is designed to offer real-time intuitive and interactive visualization to trainees. The see-through view glasses allow a trainee to visualize the radiation fields that are superimposed on the actual surrounding in the FGI suit. The Hololens, which is a distance-camera, can provide real-time visual and sound alarms to the trainee when he or she is involved in a high exposure location or posture with radiation exposure rate exceeding a pre-defined threshold.

The radiation dose information during the mock procedure is calculated using the position of the trainee and duration in that position, the pre-calculated radiation field, and the KAP specified by the user. The position information is obtained from Hololens positional tracking system which utilizes depth sensor and RGB cameras, gyroscope and accelerometer. The calculated dose can be displayed in real-time to trainee via Hololens, and be saved in the training system for future analysis.

A conversion factor is used to covert the KAP value specified by user into dose information for each exposure scenario:

$$CF = \frac{D_{simulated}}{KAP_{simulated}},$$

where $D_{simulated}$ is the dose simulated by Monte Carlo method under a specific exposure scenario and $KAP_{simulated}$ is the Kerma air product simulated by Monte Carlo method under the same exposure scenario.

RESULTS

We use different colors to represent different radiation intensities, as shown in Figure 5a. It can be seen that the intensity of radiation field varies greatly in different regions. In addition, the user can use buttons at the bottom of the training system interface to turn on or turn off the visible X-ray radiation field. When the user clicks on the Dose_Hide button during the radiation safety training, the X-ray radiation field in the training environment will disappear and when the Dose_Show button is pressed, the radiation field will become visible to the trainee. Figure 5b shows the training environment without visible X-ray radiation field.

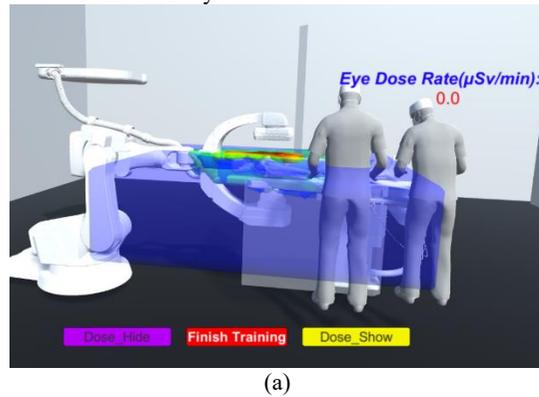

(a)

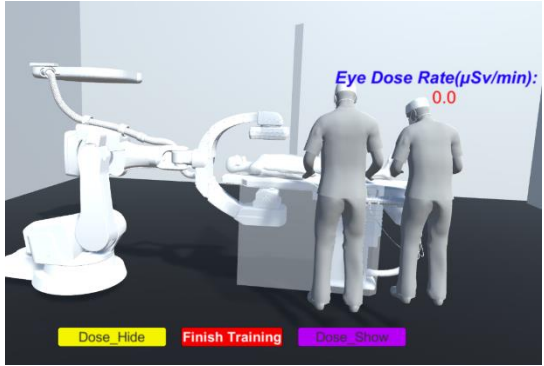

(b)

Figure 5. Training environment. (a) With X-ray radiation field. (b) Without X-ray radiation field.

Figure 6 shows the dose feedback of the radiation safety training system. The calculated results are shown in the upper-right corner of the training system interface. The total dose is the cumulative dose received by the trainee during the training session. In addition, when the dose rate exceeding a pre-defined threshold, the safety training system gives audible alarm as shown in Figure 6b.

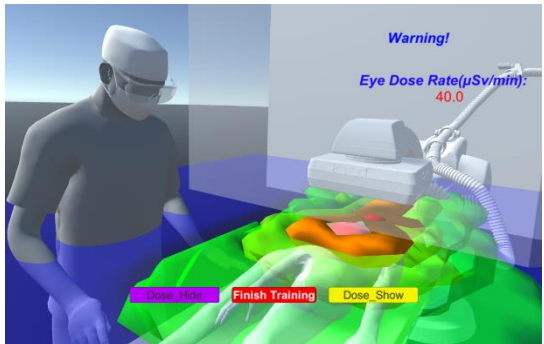

Figure 6. Dose feedback of the radiation safety training system. The trainee standing in high radiation area and the system give a warning.

Figure 7 shows the images on HoloLens glasses mapped onto a computer screen and a user manipulates the eye view by gesturing.

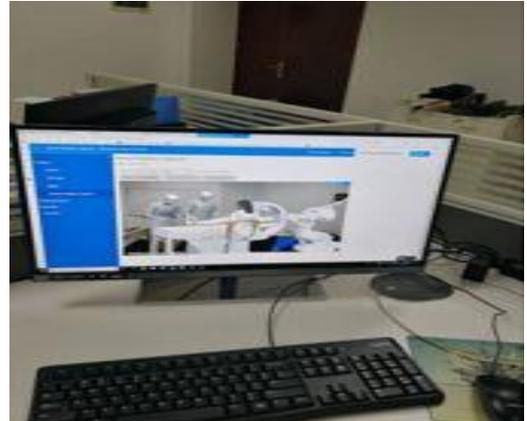

(a)

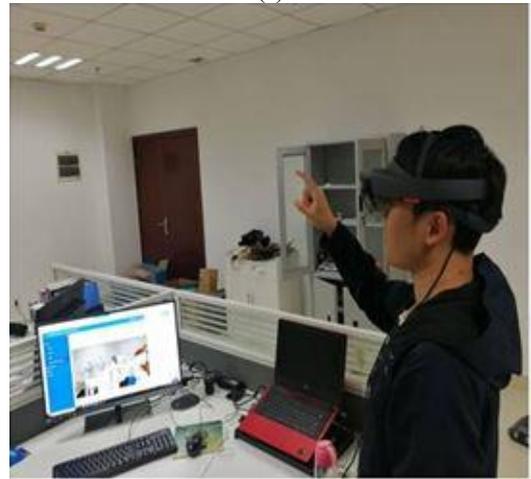

(b)

Figure 7. Functional VR system. (a) The IR suite is displayed in a computer screen. (b) The wearer of the head-mounted HoloLens system manipulates the eye view by gesturing.

DISCUSSION

To verify our computational methods, we compared staff dose information with literature data for measurement by Huda et. al [20]. In that study, Huda *et al.* used the computational framework developed at RPI to evaluate doses to several staff members in an endoscopic retrograde cholangiopancreatography (ERCP) procedure. The staff personnel doses were obtained using thermoluminescent (TL) dosemeters. To compare results with our study, we considered the similar staff and patient FGI environment which is illustrated in Figure 3b (in reference to Figure 3 in Huda. *et al.*). For a simulated training session with the same KAP rate and Fluoroscopy time as reported in the referred study, we found that the accumulated eye dose to the staff members reported by our training system was 0.02 mSv, which is in a good agreement with data reported by Huda et. al.

CONCLUSION

A preliminary VR-based radiation safety training system for operators in interventional radiology using the MS HoloLens headset device has been conceptualized. Preliminary development used the deformable whole-body computational human phantoms (RPI-Adult Male and Female) and Monte Carlo methods to calculate the X-ray field in interventional radiology. A C++ program was developed to import radiation field data, render three-dimensional models with different colors and export them to FBX format files. Differentiating the radiation intensity by different colors allowed operators to better understand the radiation field and to learn how to more effectively avoid high radiation areas during procedures. In addition, we mocked the FGI suite environment with 3D models of operators, couch, patient and X-ray source using 3D Studio Max and unity3D. The preliminary results suggest that, using MS HoloLens glasses, the training system enables the trainees to visualize the radiation field in the FGI suite intuitively and effectively. With the help of the real-time dose feedback assistance, the trainees could learn how to avoid the high radiation areas actively or adjust their postures to minimize radiation exposure in FGI procedures.

REFERENCES

1. Balter, S. Interventional fluoroscopy: physics, technology, safety. Wiley-Liss (2001).
2. Kitahara, C.M., Linet, M.S., Balter, S., et al. Occupational radiation exposure and deaths from malignant intracranial neoplasms of the brain and CNS in U.S. radiologic technologists, 1983–2012. AJR, 208:1278–1284 (2017).
3. Organization for Occupational Radiation Safety in Interventional Fluoroscopy. Invisible impact: the risk of ionizing radiation on cath lab staff. www.youtube.com/watch?v=rXgt0bF3GJM&feature=youtu.be. Accessed January 23 (2018).
4. Roguin, A., Bartal, G. Radiation and your brain. Endovascular Today, 15:63–65 (2016).
5. Miller, D.L., Vañó, E., Bartal, G., et al. Occupational radiation protection in interventional radiology: a joint guideline of the Cardiovascular and Interventional Radiology Society of Europe and the Society of Interventional Radiology. J VascIntervRadiol, 21:607–615 (2010).
6. Roguin, A., Goldstein, J., Bar, O., Goldstein, J.A. Brain and neck tumors among physicians performing interventional procedures. Am J Cardiol. May 1;111(9):1368-72 (2013).
7. Morillo, A.J. Occupational Radiation Exposure in Interventional Radiology and the Risks of Acquiring a Brain Tumor. American Journal of Roentgenology 209:6, W402-W402 (2017).
8. Kitahara, C.M., Miller, D.L. Reply to "Occupational Radiation Exposure in Interventional Radiology and the Risks of Acquiring a Brain Tumor". American Journal of Roentgenology 209:6, W403-W403 (2017).
9. National Council on Radiation Protection and Measurements. Radiation Dose Management for Fluoroscopically-guided Interventional Medical Procedures. NCRP Report No. 168 (National Council on Radiation Protection and Measurements, Bethesda, 3899 Maryland) (2010).
10. Kim, K.P., Miller, D.L., de Gonzalez, A.B., et al. Occupational radiation doses to operators performing fluoroscopically-guided procedures. Health Phys. 103(1), 80 (2012).
11. Balter, S. Stray radiation in the cardiac catheterization laboratory. Radiat. Prot. Dosim. 94(1–2), 183–188 (2001).
12. Vano, E., Gonzalez, L., Guibelalde, E., Fernandez, J.M., Ten, J.I. Radiation exposure to medical staff in interventional and cardiac radiology. Br J Radiol, 71(849):954–960 (1998).
13. Koukorava, C., Carinou, E., Simantirakis, G., Vrachliotis, T.G., Archontakis, E., Tierris, C., et al. Doses to operators during interventional radiology procedures: focus on eye lens and extremity dosimetry. RadiatProt Dosimetry, 144(1–4):482–486 (2011).
14. Stewart, F.A., Akleyev, A.V., Hauer-Jensen, M., Hendry, J.H., Kleiman, N.J., Macvittie, T.J. et al. ICRP publication 118: ICRP statement on tissue reactions and early and late effects of radiation in normal tissues and organs—threshold doses for tissue reactions in a radiation protection context. Ann ICRP 41(1–2): 1–322 (2012).
15. Dauer, L.T. Exposed medical staff: challenges, available tools, and opportunities for improvement. Health Physics, 106(2):217-224 (2014).
16. Dauer, L.T., et al. Status of NCRP Scientific Committee 1-23 Commentary on Guidance on Radiation Dose Limits for the Lens of the Eye. Health Physics, 110(2):182-184 (2016).
17. Xu, X.G. An exponential growth of computational phantom research in radiation protection, imaging, and radiotherapy: a review of the fifty-year history. Phys Med Biol. 59:R233-R302 (2014).
18. Zhang, J., Na, Y.H., Caracappa, P.F., Xu, X.G. RPI-AM and RPI-AF, a pair of mesh-based, size-adjustable adult male and female computational phantoms using ICRP-89 parameters and their calculations for organ doses from monoenergetic photon beams. Phys Med Biol. 54:5885-5908 (2009).
19. Goorley, T. MCNP6. 1.1-beta release notes. Los Alamos National Laboratory Technical Report. (2014).
20. Huda, A., Garzón, W. J., Filho, G.C.L., Vieira, B., Kramer, R., Xu, X.G., Gao, Y., and Khoury, H. J. Evaluation of staff, patient and foetal radiation doses due to endoscopic retrograde cholangiopancreatography (ERCP) procedures in a pregnant patient. Radiation protection dosimetry 168(3):401-407(2015).